\documentclass[final]{aipproc}
\layoutstyle{6x9}
\usepackage{amsmath,wrapft,epsfig}
\begin{document}

%%%%%%%%%%%%%%%%%%%%%%%%%%%%%%%%%%%%%%%%%%%%%%%%%%%%%%%%%%%%%%%%%%%%%%%%%%%%%%%%
%%%%%%%%%%%%%%%%%%%%%%%%%%%%%%%%%%%%%%%%%%%%%%%%%%%%%%%%%%%%%%%%%%%%%%%%%%%%%%%%
\title{Determination of fragmentation functions and 
       their uncertainties from $e^+ + e^- \rightarrow h + X$ data}


\author{M. Hirai}
{address={Department of Physics,
             Tokyo Institute of Technology \\
             Ookayama, Meguro-ku, Tokyo, 152-8550, Japan}}
\author{S. Kumano}
{address={Institute of Particle and Nuclear Studies,
          High Energy Accelerator Research Organization (KEK) \\
          1-1, Ooho, Tsukuba, Ibaraki, 305-0801, Japan}
,altaddress={Department of Particle and Nuclear Studies,
             Graduate University for Advanced Studies \\
           1-1, Ooho, Tsukuba, Ibaraki, 305-0801, Japan}}
\author{T.-H. Nagai}
{address={Department of Particle and Nuclear Studies,
             Graduate University for Advanced Studies \\
           1-1, Ooho, Tsukuba, Ibaraki, 305-0801, Japan}}
\author{K. Sudoh}
{address={Institute of Particle and Nuclear Studies,
          High Energy Accelerator Research Organization (KEK) \\
          1-1, Ooho, Tsukuba, Ibaraki, 305-0801, Japan}}

\begin{abstract}
Fragmentation functions are determined for pions, kaons, and nucleons
by a global analysis of charged-hadron production data in electron-positron
annihilation. The optimum functions are obtained in both leading order (LO)
and next-to-leading order (NLO) of $\alpha_s$. It is important that
uncertainties of the fragmentation functions are estimated in this work
by the Hessian method. We found that the uncertainties are large
at small $Q^2$ and that they are generally reduced in the NLO in comparison
with the LO ones. We supply a code for calculating the fragmentation
functions and their uncertainties for the pions, kaons, and nucleons
at given $z$ and $Q^2$.
\end{abstract}

\maketitle
%%%%%%%%%%%%%%%%%%%%%%%%%%%%%%%%%%%%%%%%%%%%%%%%%%%%%%%%%%%%%%%%%%%%%%%%%%%%%%%%
%%%%%%%%%%%%%%%%%%%%%%%%%%%%%%%%%%%%%%%%%%%%%%%%%%%%%%%%%%%%%%%%%%%%%%%%%%%%%%%%
\section{Introduction}

Fragmentation functions are used for calculating cross sections of
high-energy hadron-production processes. They describe how a hadron
is produced in the final state from a parent quark, antiquark, or gluon.
Recently, they are becoming increasingly important because semi-inclusive
hadron-production processes are investigated for finding the origin
of nucleon spin and properties of quark-hadron matters in lepton-nucleon
scattering and hadron-hadron collisions.

There have been studies of determining the functions from hadron-production
data in electron-positron annihilation \cite{ffs-studies}. It is known
that determined functions are very different between the parametrization
groups, for example KKP (Kniehl, Kramer, and Potter) and Kretzer.
It is unfortunate that uncertainties of the functions were not estimated
although they have been studied in parton distribution functions
(PDFs) in the nucleon \cite{pdf-error-n,pdf-error-pol} and nuclei
\cite{pdf-error-a}. The major purpose of our work is to investigate
the uncertainties of the fragmentation functions \cite{ffs06}
in both leading order (LO) and next-to-leading order (NLO) of
the running coupling constant $\alpha_s$.

%%%%%%%%%%%%%%%%%%%%%%%%%%%%%%%%%%%%%%%%%%%%%%%%%%%%%%%%%%%%%%%%%%%%%%%%%%%%%%%%
%%%%%%%%%%%%%%%%%%%%%%%%%%%%%%%%%%%%%%%%%%%%%%%%%%%%%%%%%%%%%%%%%%%%%%%%%%%%%%%%
\section{Analysis method}

The fragmentation functions are determined by analyzing the data
for charged-hadron production in electron-positron annihilation.
The fragmentation function for hadron $h$ is defined by the cross section
for $e^+ +e^- \rightarrow h+X$:
\begin{equation}  
F^h(z,Q^2) = \frac{1}{\sigma_{tot}} 
\frac{d\sigma (e^+e^- \rightarrow hX)}{dz} ,
\label{eqn:def-ff}
\end{equation}
where $Q^2=s$ with the center-of-mass energy $\sqrt{s}$, and
$\sigma_{tot}$ is the total hadronic cross section. The variable $z$
is given by the hadron energy $E_h$ and the beam energy $\sqrt{s}/2$ by
$z \equiv E_h / (\sqrt{s}/2) = 2E_h /Q$.
The fragmentation process occurs from primary quarks, antiquarks,
and gluons, so that the function is expressed by their contributions:
$F^h(z,Q^2) = \sum_i C_i(z,\alpha_s) \otimes D_i^h (z,Q^2)$,
where $D_i^h(z,Q^2)$ is a fragmentation function of the hadron $h$ 
from a parton $i$ ($=u,\ d,\ s,\ \cdot\cdot\cdot,\ g$),
$C_i(z,\alpha_s)$ is a coefficient function, and
$\otimes$ indicates a convolution integral \cite{ffs06}.

In our analysis, the functions are defined at fixed $Q^2$ ($\equiv Q_0^2$)
as \cite{ffs06}:
\begin{equation}
D_i^h(z,Q_0^2) = N_i^h z^{\alpha_i^h} (1-z)^{\beta_i^h} ,
\end{equation}
where $N_i^h$, $\alpha_i^h$, and $\beta_i^h$ are parameters.
Instead of $N_i^h$, we use a second moment $M_i^h$ as a parameter
in our analysis because its physical meaning is clearer. They
are related by
$ N_i^h = M_i^h / B(\alpha_i^h+2, \beta_i^h+1) $,
where $B(\alpha_i^h+2, \beta_i^h+1)$ is the beta function.
The parameters are determined from a $\chi^2$ analysis of the data.
The data are taken at large $Q^2$, typically $Q^2 = M_Z^2$, and the functions
are evolved to experimental $Q^2$ points by the timelike DGLAP equations.
The parameters are determined by minimizing the total $\chi^2$,
$\chi^2 = \sum_j (F_{j}^{data}-F_{j}^{theo})^2 / (\sigma_j^{data})^2$,
where $F_{j}^{data}$ and $F_{j}^{theo}$ are experimental and
theoretical fragmentation functions, and $\sigma_j^{data}$ is
an experimental error. The uncertainties of the obtained
functions are calculated by the Hessian method
\cite{pdf-error-n,pdf-error-pol,pdf-error-a,ffs06}:
\begin{equation}
[\delta D_i^h (z)]^2=\Delta \chi^2 \sum_{j,k}
\left( \frac{\partial D_i^h (z,\xi)}{\partial \xi_j}  \right)_{\hat\xi}
H_{jk}^{-1}
\left( \frac{\partial D_i^h (z,\xi)}{\partial \xi_k}  \right)_{\hat\xi}
\, ,
\end{equation}
where $H_{ij}$ is the Hessian, $\xi_i$ is a parameter,
and the parameter set at the minimum $\chi^2$ is denoted by $\hat \xi$.

%%%%%%%%%%%%%%%%%%%%%%%%%%%%%%%%%%%%%%%%%%%%%%%%%%%%%%%%%%%%%%%%%%%%%%%%%%%%%%%%
%%%%%%%%%%%%%%%%%%%%%%%%%%%%%%%%%%%%%%%%%%%%%%%%%%%%%%%%%%%%%%%%%%%%%%%%%%%%%%%%
\section{Results}

The parameters are determined by fitting the data for
$e^+e^- \rightarrow hX$. As an example, the obtained fragmentation
functions in the NLO are compared with experimental data for
the charged pions ($\pi^+ + \pi^-$) in Fig. 1,
where fractional differences $(Data-Theory)/Theory$ are shown for
the fragmentation functions at the experimental $Q^2$ points.
There are fourteen parameters in the pion analysis. The number
of the data is 264. The bands indicate the one-$\sigma$ uncertainty
ranges estimated by the Hessian method. The data are generally well
explained by our parametrization; however, the DELPHI data
significantly deviate from our fit.
The total $\chi^2$ is 433.5 in the NLO analysis for the pion
so that $\chi^2$/d.o.f. is 1.73. In the LO analysis, the total $\chi^2$
is slightly larger.

Determined fragmentation functions for $\pi^+$ are shown 
at $Q^2$=1 GeV$^2$, $m_c^2$, $m_b^2$, and $M_Z^2$ in Fig. 2.
The functions in the LO and NLO are indicated by the dashed
and solid curves, and the uncertainties in the LO and NLO are shown 
by the light- and dark-shaded bands, respectively. It is important
that the uncertainties are now shown in this work for
the fragmentation functions.
At small $Q^2$ ($\sim 1$ GeV$^2$), the uncertainties are large,
especially in the LO. The uncertainty bands become smaller
in the NLO. It is very difficult to determine the gluon function;
however, an improvement can be seen in the NLO because the uncertainty
becomes much smaller. The functions are relatively well determined
at large $Q^2$ ($\sim M_Z^2$), where the $e^+e^-$ data are taken.

\vspace{-0.25cm}
%%%%%%%%%%%%%%%%%%%%%%%%%%%%%%%% figure %%%%%%%%%%%%%%%%%%%%%%%%%%%%%%%%%%%%%%
\noindent
\parbox[t]{0.46\textwidth}{
   \begin{center}
       \epsfig{file=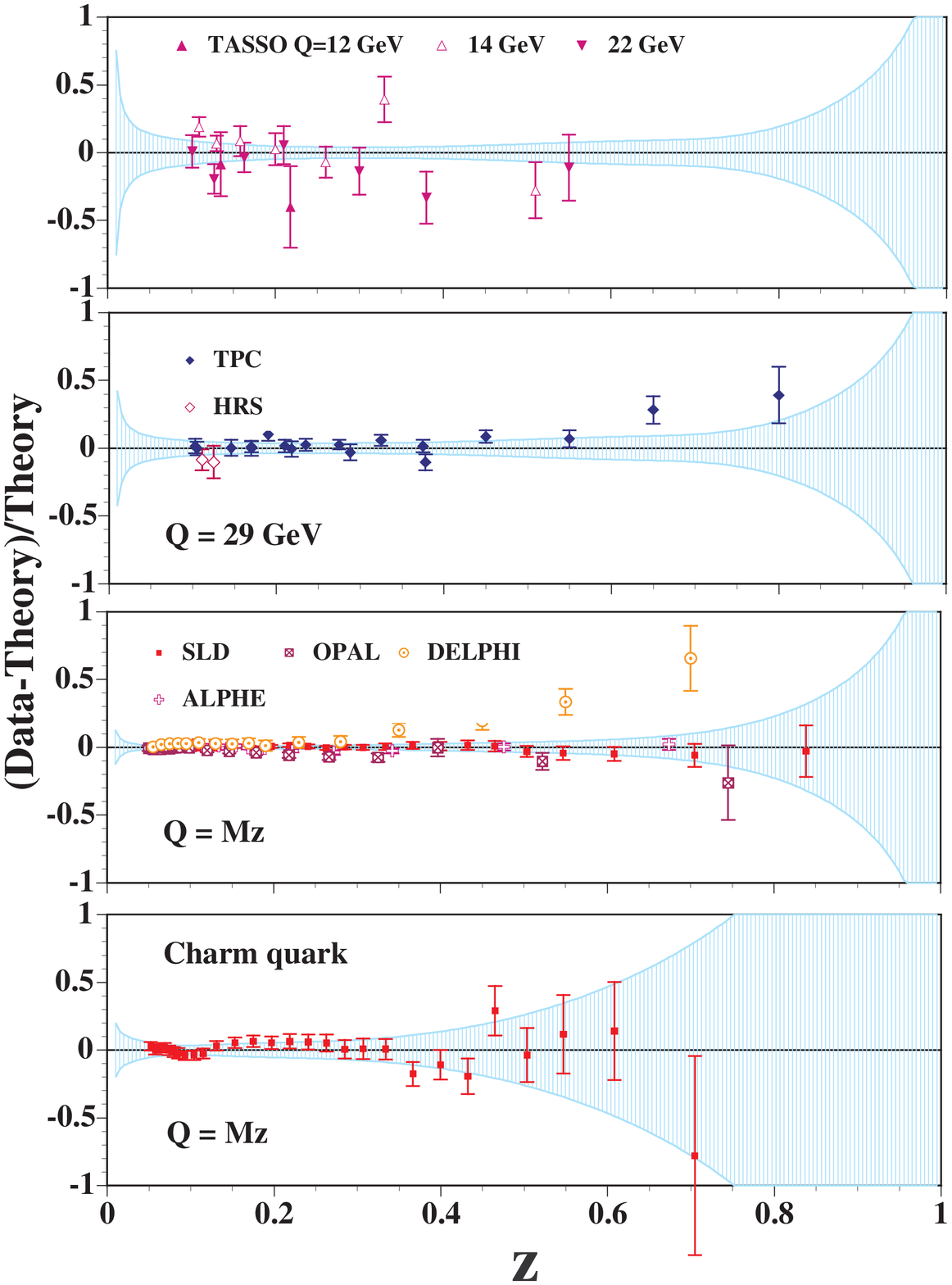,width=5.8cm} \\
   \end{center}
}\hfill
\parbox[t]{0.46\textwidth}{
   \begin{center}
       \vspace{-0.1cm}
       \epsfig{file=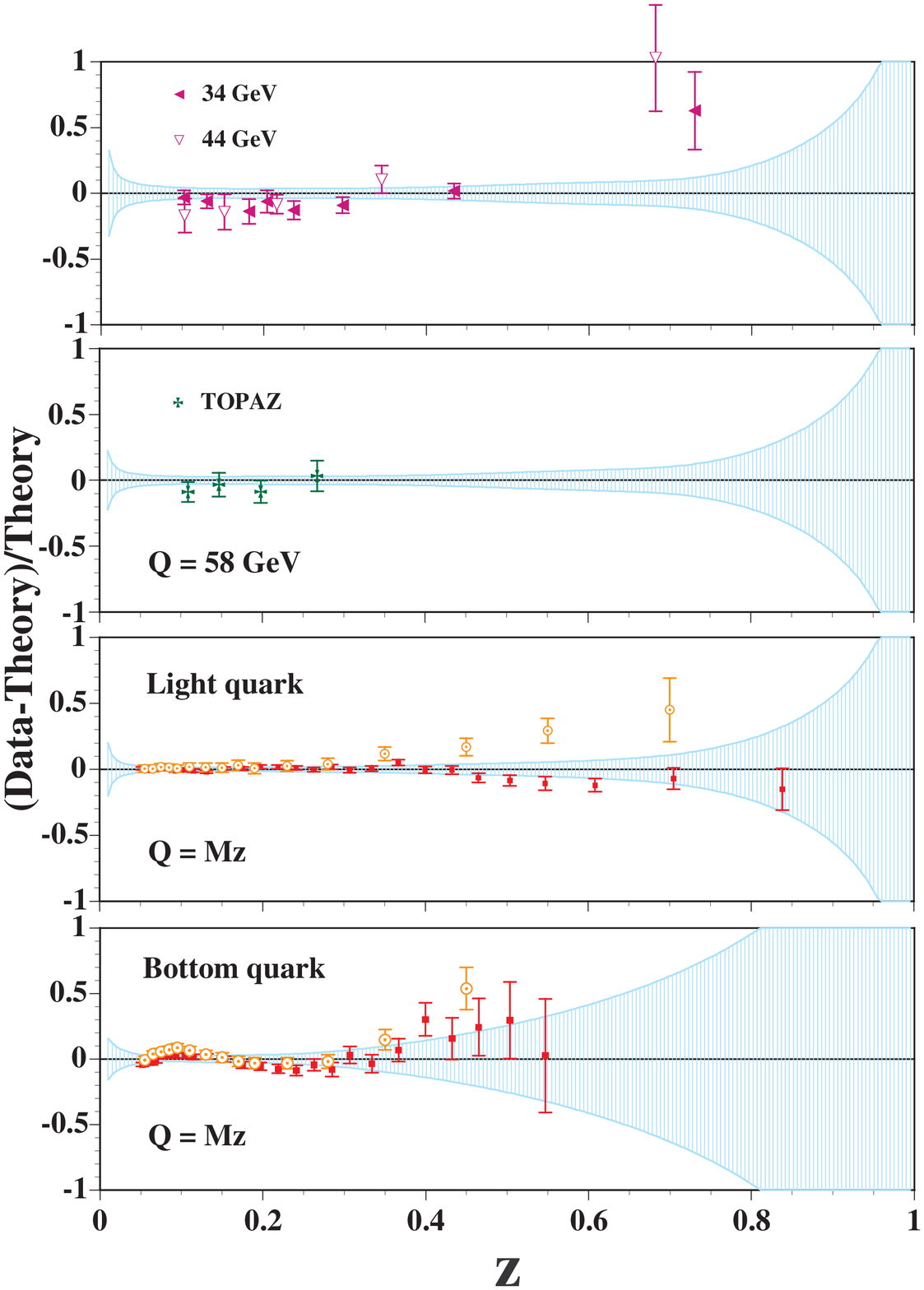,width=5.8cm} \\
   \end{center}
}
   \begin{center}
\vspace{-0.3cm}
       {\footnotesize {\bf FIGURE 1.}
          Comparison with experimental data for pions \cite{ffs06}.}
   \end{center}
%%%%%%%%%%%%%%%%%%%%%%%%%%%%%%%% figure %%%%%%%%%%%%%%%%%%%%%%%%%%%%%%%%%%%%%%
\vspace{-0.25cm}
%%%%%%%%%%%%%%%%%%%%%%%%%%%%%%%% figure %%%%%%%%%%%%%%%%%%%%%%%%%%%%%%%%%%%%%%
\noindent
\parbox[t]{0.46\textwidth}{
   \begin{center}
       \epsfig{file=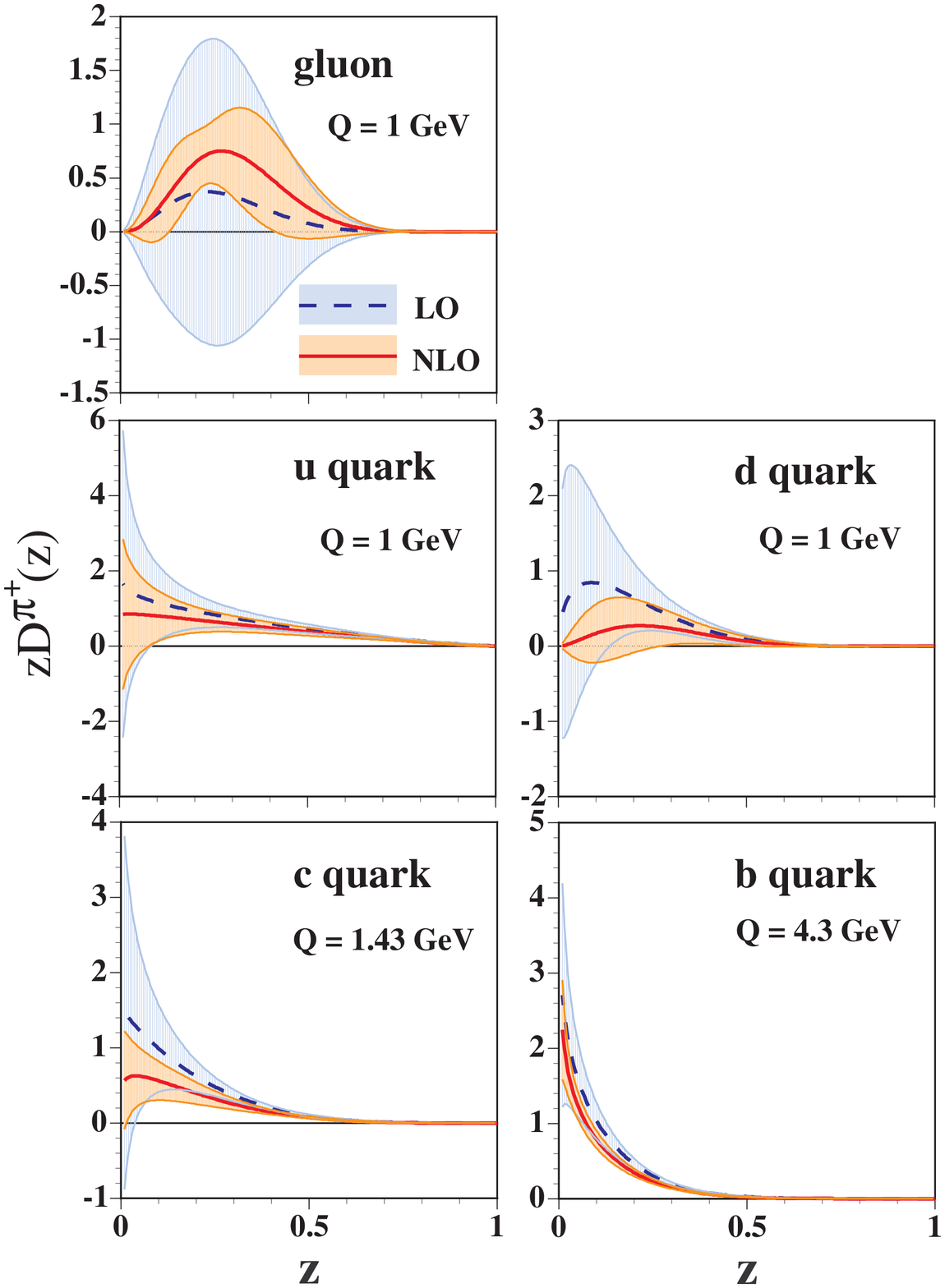,width=5.8cm} \\
   \end{center}
}\hfill
\parbox[t]{0.46\textwidth}{
   \begin{center}
       \vspace{-0.1cm}
       \epsfig{file=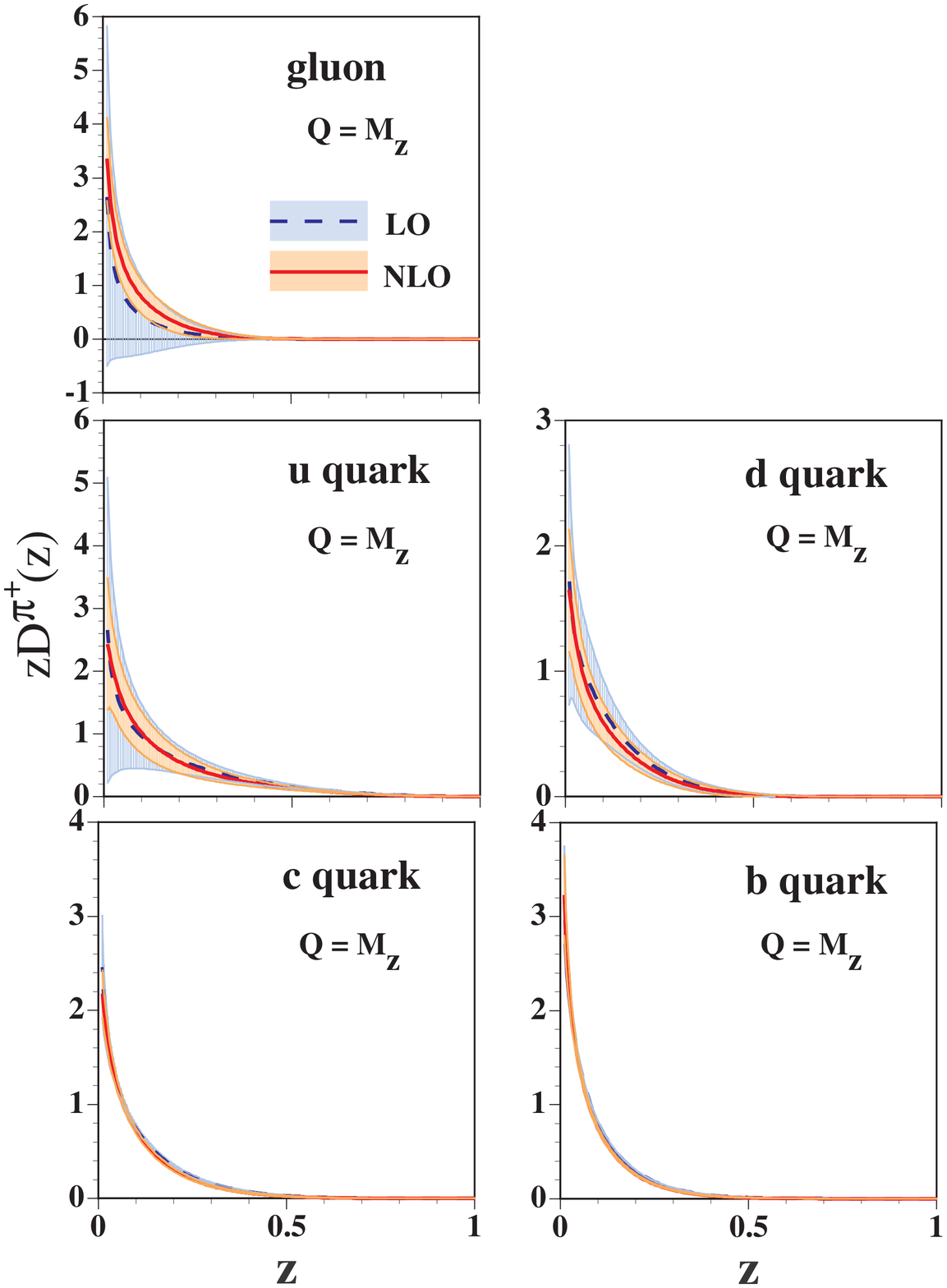,width=5.8cm} \\
   \end{center}
}
   \begin{center}
\vspace{-0.3cm}
       {\footnotesize {\bf FIGURE 2.}
          Determined fragmentation functions for $\pi^+$ at $Q^2=1$ GeV$^2$,
             $m_c^2$, $m_b^2$, and $M_Z^2$ \cite{ffs06}.}
   \end{center}
%%%%%%%%%%%%%%%%%%%%%%%%%%%%%%%% figure %%%%%%%%%%%%%%%%%%%%%%%%%%%%%%%%%%%%%%
\vspace{0.0cm}

\vspace{0.0cm}
%%%%%%%%%%%%%%%%%%%%%%% figure %%%%%%%%%%%%%%%%%%%%%%%
\begin{wrapfigure}{r}{0.48\textwidth}
   \vspace{-0.3cm}
   \begin{center}
       \epsfig{file=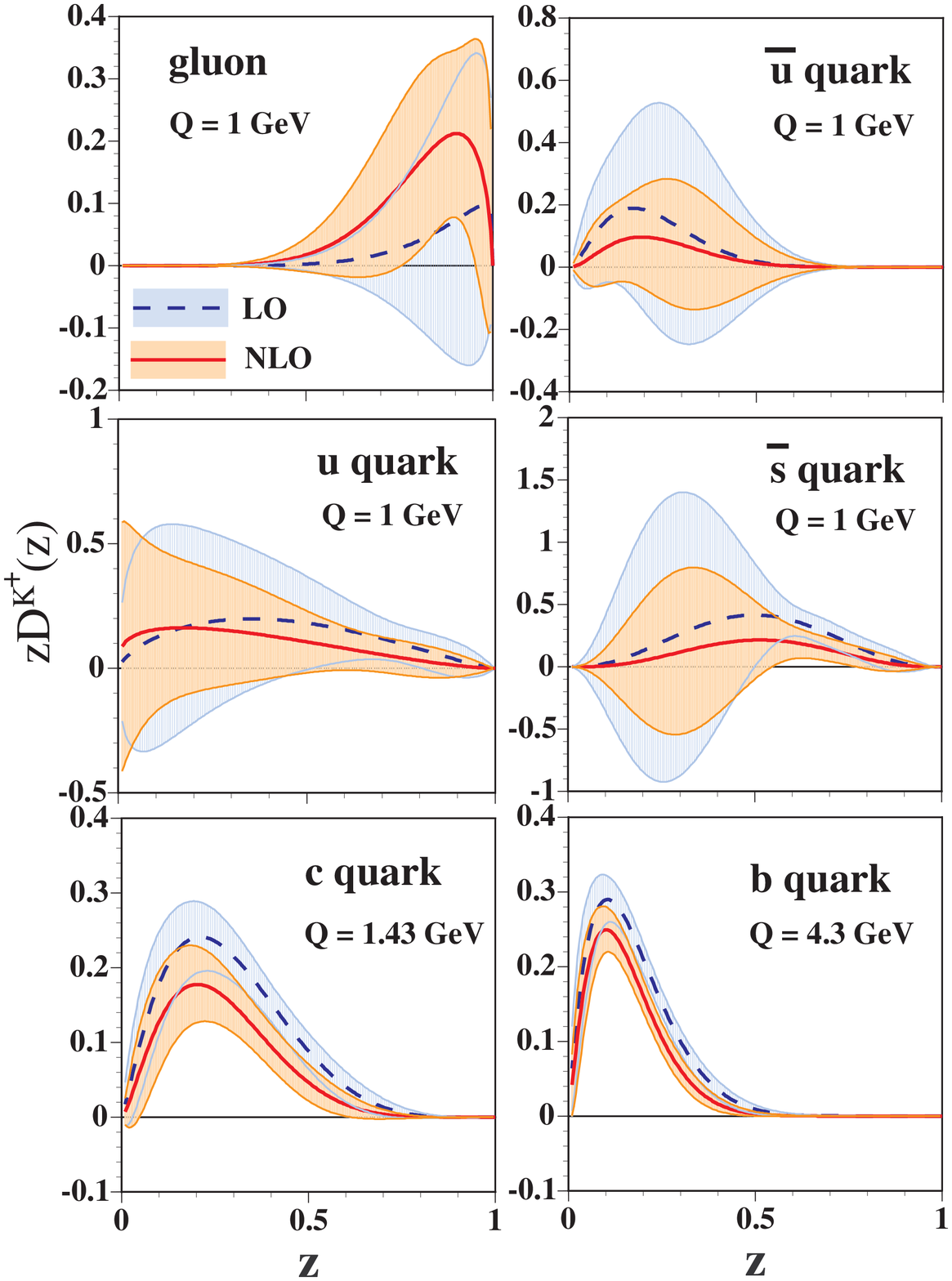,width=5.8cm} \\
       \vspace{0.1cm}
       {\footnotesize \hspace{0.15cm} {\bf FIGURE 3.}
            Fragmentation functions for $K^+$ \cite{ffs06}.}
   \end{center}
   \vspace{-0.3cm}
\end{wrapfigure}
%%%%%%%%%%%%%%%%%%%%%%% figure %%%%%%%%%%%%%%%%%%%%%%%
\vspace{0.0cm}

Similar analyses have been done for the kaons and proton/anti-proton.
In both cases, the values of $\chi^2$/d.o.f. are about two, and
the data are generally well explained by our parametrization.
As an example, the functions for $K^+$ are shown in Fig. 3 with
their uncertainty bands. The notations are the same as the ones
in Fig. 2. The functions have large uncertainty bands also for
the kaon at small $Q^2$.
The large uncertainties suggest that the hadron-production
cross sections have large ambiguities if they are calculated in the small
$p_T$ region. The obtained functions for the proton as well as
the detailed discussions on the pions and kaons are found in
Ref. \cite{ffs06}.

From the $\chi^2$ analysis of the data for $e^+e^-$$\rightarrow hX$,
the optimum fragmentation functions are determined for the pions,
kaons, and nucleons. It is especially important that the uncertainties
of the obtained functions are calculated in this work. The results 
indicate large uncertainties at small $Q^2$, which suggests large
ambiguities in extracting information, for example on nucleon spin
and quark-hadron matters from hadron-production cross sections in
lepton-hadron and hadron-hadron reactions. In such studies, 
it is important to indicate reliable regions of the fragmentation
functions by using the uncertainties. 
We supply our code for calculating not only the optimum
fragmentation functions but also their uncertainties at given
$z$ and $Q^2$ for general users \cite{ffs06}.

\vspace{-0.1cm}
%%%%%%%%%%%%%%%%%%%%%%%%%%%%%%%%%%%%%%%%%%%%%%%%%%%%%%%%%%%%%%%%%%%%%%%%%%%%%%%%
%%%%%%%%%%%%%%%%%%%%%%%%%%%%%%%%%%%%%%%%%%%%%%%%%%%%%%%%%%%%%%%%%%%%%%%%%%%%%%%%
\begin{theacknowledgments}
\vspace{-0.20cm}
S.K. and M.H. were supported by the Grant-in-Aid for Scientific Research from
the Japanese Ministry of Education, Culture, Sports, Science, and Technology.
T.-H.N. was supported by the JSPS Research Fellowships for Young Scientists.
\end{theacknowledgments}

\vspace{-0.1cm}
%%%%%%%%%%%%%%%%%%%%%%%%%%%%%%%%%%%%%%%%%%%%%%%%%%%%%%%%%%%%%%%%%%%%%%%%%%%%%%%%
%%%%%%%%%%%%%%%%%%%%%%%%%%%%%%%%%%%%%%%%%%%%%%%%%%%%%%%%%%%%%%%%%%%%%%%%%%%%%%%%
\bibliographystyle{aipprocl}

\end{document}